# High-Throughput Computational-Experimental Screening Protocol for the Discovery of Bimetallic Catalysts

Byung Chul Yeo,[1,2,†] Hyunji Nam,[3,4,†] Hyobin Nam,[3,5] Min-Cheol Kim,[1] Hong Woo Lee,[1,4] Donghun Kim,[1] Kwan-Young Lee,[4] Seung Yong Lee,[3,5,]* and Sang Soo Han[1,]*

[1]Computational Science Research Center, Korea Institute of Science and Technology, Seoul 02792, Republic of Korea

[2]Department of Energy Resources Engineering, Pukyong National University, Busan 48513, Republic of Korea

[3]Materials Achitecturing Research Center, Korea Institute of Science and Technology, Seoul 02792, Republic of Korea

[4]Department of Chemical and Biological Engineering, Korea University, Seoul 02841, Republic of Korea

[5]Department of Nanomaterials Science and Engineering, Korea University of Science and Technology, Daejeon 34113, Republic of Korea

*Correspondence to: sangsoo@kist.re.kr (S.S.H); patra@kist.re.kr (S.Y.L)

[†]These authors contributed equally to this work.

## Abstract

For decades of catalysis research, the *d*-band center theory that correlates the *d*-band center and the adsorbate binding energy has successfully enabled the accelerated discovery of novel catalyst materials. Recent studies indicate that, on top of the *d*-band center value, the full consideration of the *d*-band shapes describing higher moments of the *d*-band as well as *sp*-band properties can help better capturing surface reactivity. However, the density-of-states (DOS) patterns themselves have never been used as a descriptor in combined computational-experimental studies. Here, we propose the full DOS patterns as a key descriptor in high-throughput screening protocols, and prove its effectiveness. For the hydrogen peroxide ($H_2O_2$) synthesis as our demo catalytic reaction, the present study focuses on discovering bimetallic catalysts that can replace the prototypic palladium (Pd) one. Through a series of screening processes based on DOS pattern similarities (evaluated using first-principles calculations) and synthetic feasibility, 9 candidates are finally proposed out of 4,350 bimetallic alloy, which then are expected to have a catalytic performance comparable to that of Pd. The subsequent experimental tests demonstrate that 4 bimetallic catalysts ($Ni_{61}Pt_{39}$, $Au_{51}Pd_{49}$, $Pt_{52}Pd_{48}$, $Pd_{52}Ni_{48}$) indeed exhibit the catalytic properties comparable to those of Pd. Moreover, we discovered a novel bimetallic (Ni-Pt) catalyst, which has not yet been reported for $H_2O_2$ direct synthesis. In particular, $Ni_{61}Pt_{39}$ outperforms the prototypical Pd catalyst for the chemical reaction and exhibits a 9.5-fold enhancement in cost-normalized productivity. This protocol provides a new opportunity for the catalyst discovery for the replacement or reduction in use of the platinum-group metals.

## Introduction

One of the important roles of computer simulations in materials science is to predict and/or design novel materials prior to experiments. In particular, first-principles calculations using density functional theory (DFT) have played a vital role in the field of catalysis.[1-3] The computational framework enables the prediction or understanding of reaction pathways on catalyst surfaces, which is undoubtedly useful for experimentalists. However, the accurate estimation of some crucial properties such as the reaction barriers on catalytic surfaces is extremely time-consuming in first-principles calculation scheme. In this regard, obtaining a full understanding of the catalytic reaction mechanism via first-principles calculations is often considered rather inefficient. In our experiences, though computing-power-dependent, it requires time costs of several months even for computing a full energetics picture of one metallic catalyst system. The efficiency has therefore been questioned for the first-principles-calculation-guided search of novel catalysts, because carrying out experimental tests without the aid of computer simulations could often be even faster.

To maximize the efficiency of novel catalyst discovery using the combined computational and experimental study, the descriptors or features that bridge these two area should be wisely chosen. To be specific, a simple but physically reasonable descriptor or feature enabling the representation of catalytic properties is critical. A representative example of such a descriptor is the *d*-band center theory that correlates between the *d* band center,[4] i.e., the average energy of electronic *d* states projected onto a surface atom of the catalyst, and the gas adsorption energy, which has been widely accepted in the field of metallic catalysis. Coupled with the volcano relationship between catalytic activity and adsorption energy,[5] it is theoretically possible to predict catalytic activity from the *d*-band center without exploration of the full reaction mechanism. Afterward, along with the *d*-band center theory, consideration of the *d*-band shapes describing higher moments of the *d*-band as well as the *sp*-band properties describing local Pauli electronegativity were also suggested to capture the surface reactivity of transition metal alloys.[6-8] These studies indicate that the electronic density of states (DOS) patterns themselves projected onto surface atoms of catalysts can serve as an improved descriptor for the development of novel metal catalysts because they include more comprehensive information of not only *d* states but also *sp* states (both their values and shapes). Nonetheless, the full DOS patterns themselves have never been used as a descriptor in combined computational-experimental screening process.

Since electronic structures are key to determining the physical/chemical properties of materials, materials with similar electronic structures tend to exhibit the similar properties. Indeed, a DFT study revealed that a solid-solution $Ir_{50}Au_{50}$ alloy can exhibit catalytic properties for $H_2$ dissociation reaction with a comparable activity of Pt, and this result is attributed to their similar electronic structures (DOS patterns) between the $Ir_{50}Au_{50}$ alloy and Pt.[9] In another example of $Rh_{50}Ag_{50}$ versus Pd for hydrogen storage, their similar electronic structures were also experimentally verified.[10] The similar electronic structural characteristics enables the $Rh_{50}Ag_{50}$ alloy to exhibit hydrogen storage properties as superior as Pd,[11] although neither Rh nor Ag alone shows hydrogen storage properties.[9] Thus, bimetallic alloys with a DOS pattern similar to that of Pd, which is well-known as a representative metal catalyst, would exhibit catalytic performance similar to that of Pd, which is a good hypothesis for the high-throughput development of bimetallic catalysts.

Here, we use the full DOS pattern as a key descriptor in the high-throughput computational-experimental screening protocols. As an exemplary demonstration in the high-throughput search of bimetallic catalysts, we select palladium (Pd), the prototypical catalyst for hydrogen peroxide ($H_2O_2$) synthesis from hydrogen ($H_2$) and oxygen ($O_2$) gases, [12-14] as our reference material. Using DFT calculations, we first screened 4,350 crystal structures of bimetallic alloys to investigate their thermodynamic stabilities (only closest-packed surfaces considered). Next, the similarities in the DOS patterns between the alloys and Pd were quantified, and subsequently their synthetic feasibility was also evaluated. Through these processes, we finally proposed 9 candidates with the high DOS similarities with Pd, which then are expected to show catalytic properties comparable to those of Pd. These 9 screened alloys were experimentally synthesized and tested for $H_2O_2$ direct synthesis, and 4 of them ($Ni_{61}Pt_{39}$, $Au_{51}Pd_{49}$, $Pt_{52}Pd_{48}$, $Pd_{52}Ni_{48}$) indeed exhibit catalytic properties comparable to that of Pd. In particular, the Pd-free $Ni_{61}Pt_{39}$ is novel (no report available in $H_2O_2$ synthesis) and found much to outperform the prototypic Pd with a 9.5-fold enhancements in cost-normalized productivity owing to the high content of inexpensive Ni.

## Results

### High-throughput computational screening for novel bimetallic catalysts

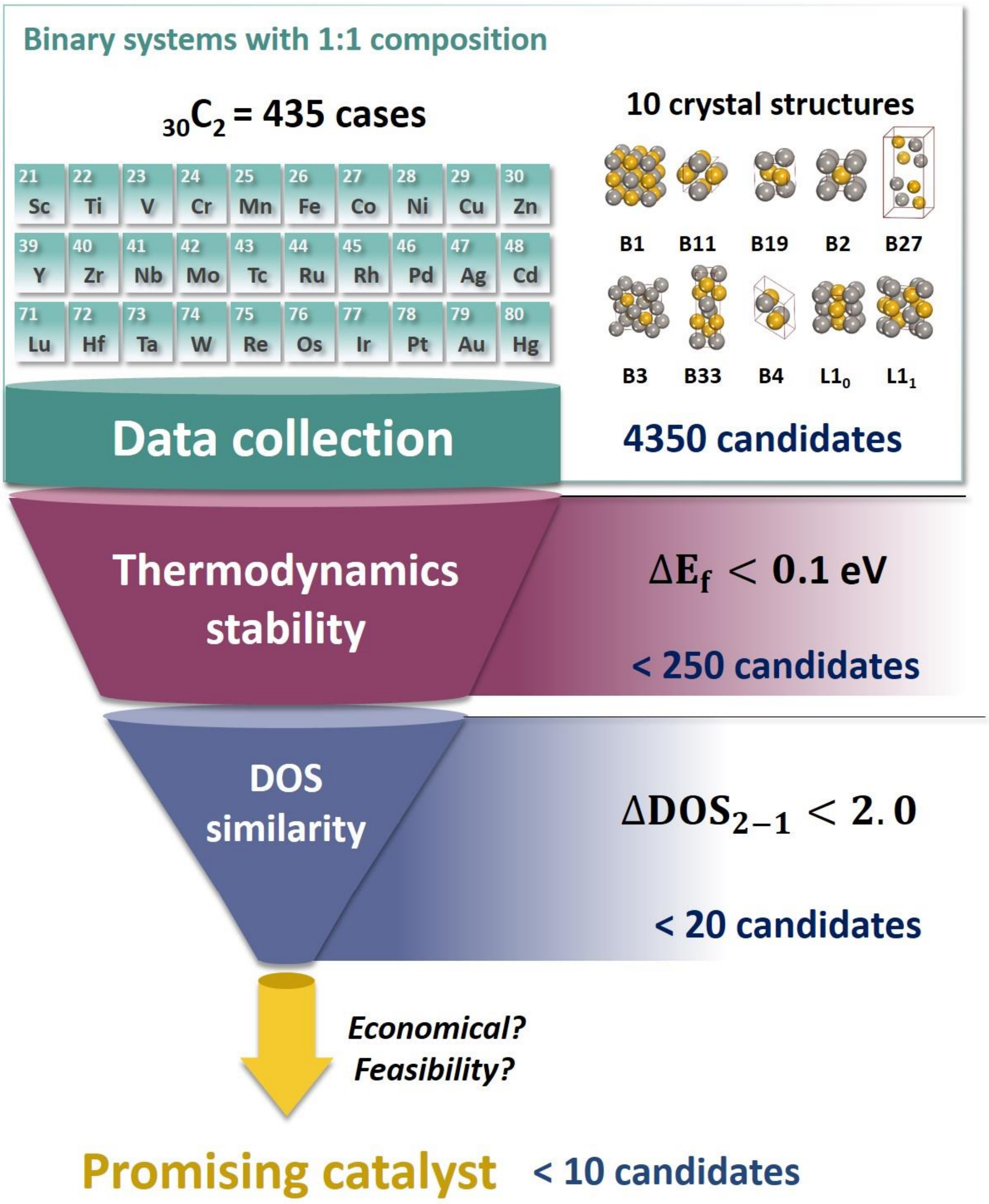

**Figure 1. High-throughput screening protocol for the discovery of bimetallic catalysts.** There are three screening steps to identify catalyst candidates from 4,350 candidates. The top (cyan color) section indicates the first step showing 30 elements and 10 crystal structures to collect binary alloys with an 1:1 composition. The middle (magenta color) and bottom (violet color) sections indicate the second and the last steps relevant to the thermodynamic stability and the DOS similarity, respectively.

Fig. 1 shows a schematic diagram of our high-throughput screening protocol for the discovery of bimetallic catalysts. Based on 30 transition metals in periods IV, V, and VI, we considered a total of 435 ($_{30}C_2$) binary systems with a 1:1 (50:50) composition. For each alloy combination, 10 ordered phases that are available for the 1:1 composition were investigated, namely, B1, B2, B3, B4, B11, B19, B27, B33, $L1_0$, and $L1_1$, leading to a screening of 4,350 (435×10) crystal structures. Using DFT calculations, we calculated the formation energy ($\Delta E_f$) of each phase and determined which crystal structure is most stable in the bulk phase. Here, a negative formation energy implies that the phase is thermodynamically favorable (or miscible element combination), and *vice versa*. Because nonequilibrium alloyed phases can be stabilized by the nanosize effect, although the elemental combinations are immiscible in the bulk phases,[11,15,16] a margin of $\Delta E_f < 0.1$ eV was considered when screening the thermodynamic stabilities of the bimetallic systems. Eventually, of a total of 435 binary systems, 249 alloys were filtered at the thermodynamic screening step.

For the 249 thermodynamically screened alloys, we calculated the DOS pattern projected on the close-packed surface for each structure and compared it with that of the Pd(111) surface. To quantitatively compare the similarity of two DOS patterns, we defined the following measure:

$$\Delta \mathrm{DOS}_{2-1} = \left\{ \int [\mathrm{DOS}_2(\mathrm{E}) - \mathrm{DOS}_1(\mathrm{E})]^2 g(\mathrm{E};\sigma) d\mathrm{E} \right\}^{\frac{1}{2}}$$

$$g(\mathrm{E};\sigma) = \frac{1}{\sigma\sqrt{2\pi}} e^{-\frac{(\mathrm{E}-\mathrm{E_F})^2}{2\sigma^2}} \quad (1)$$

Here, $\Delta \mathrm{DOS}_{2-1}$ indicates the similarity between the DOS of an alloy ($DOS_2$) and the DOS of the reference Pd (111) ($DOS_1$) with a higher weight near the Fermi energy ($E_F$) within the range of $\sigma$ = 7 eV.

The DOS similarity values of the 249 alloy surfaces are shown in Fig. 2. According to the definition of the DOS similarity, as the value approaches zero, the electronic structure of the alloy becomes more similar to that of Pd(111), and the material is expected to show catalytic properties similar to those of Pd(111). In Fig. 2, we chose 17 candidates with the low DOS similarity values ($\Delta \mathrm{DOS}_{2-1} < 2.0$): CrRh ($\Delta \mathrm{DOS}_{2-1} = 1.97$, B2 structure), FeCo (1.63, B2), CoNi (1.71, $L1_0$), CoHf (1.89, $L1_0$), CoTa (1.96, $L1_0$), CoPt (1.78, $L1_0$), NiCu (1.11, $L1_0$), NiNb (1.96, $L1_0$), NiMo (1.87, $L1_0$), NiRh (0.98, $L1_0$), NiPd (0.84, $L1_0$), NiIr (1.28, $L1_0$), NiPt (1.16, $L1_0$), CuPd (1.51, B2), RhPt (1.67, $L1_0$), PdPt (1.16, $L1_0$), and PdAu (1.43, $L1_0$).

Here, due to the rarity of Hf and Ta, the CoHf and CoTa alloys were excluded from the candidate list. Additionally, we excluded NiCu and CrRh from the list because the large reduction potential difference between the elements in the binary system leads to unfavorable formation of the alloyed structures in which two elements are homogeneously mixed via a wet-chemical method for nanoparticle (NP) synthesis.[17]

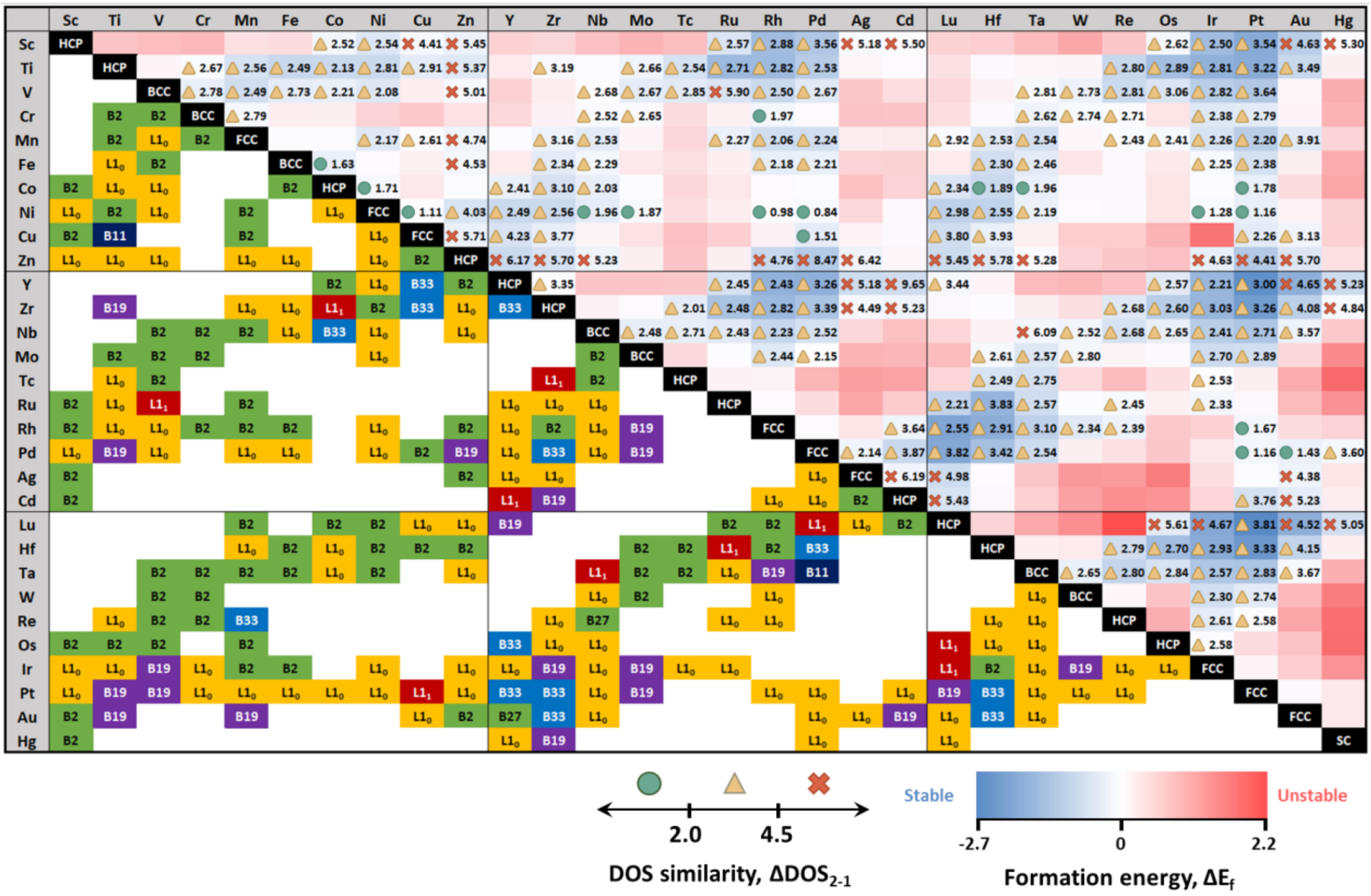

**Figure 2. Map of the thermodynamic stabilities and DOS similarities of bimetallic alloy systems.** The diagonal components denote the most thermodynamically stable crystal structures (e.g., FCC, BCC, HCP, and SC) of pure metals. The occupied components under the diagonal denote the most thermodynamically stable crystal structures (e.g., B2, B11, B19, B27, B33, $L1_0$, and $L1_1$) of bimetallic alloys. The background colors above the diagonal describe the formation energies, and the occupied components above the diagonal denote the DOS similarity of each alloy compared with Pd (111).

## Experimental verification for the DFT-guided bimetallic catalysts

For the remaining 13 candidates, we tried to synthesize their alloyed nanoparticles (NPs) and confirm whether their crystal structures are identical to those (Fig. 2) predicted by DFT calculations. For NP synthesis, we used a butyllithium reduction method.[18] Details of the NP synthesis are found in the Methods section. Because it was assumed in Fig. 2 that the proposed alloys had a composition of 50:50, we intended to synthesize alloys with nearly the same composition. Inductively coupled plasma (ICP) analyses revealed that the compositions of the synthesized alloys (the loading ratio of metal precursors = 50:50) were $Au_{51}Pd_{49}$, $Ni_{54}Co_{46}$, $Ni_{61}Pt_{39}$, $Pd_{50}Cu_{50}$, $Pd_{52}Ni_{48}$, $Pt_{58}Co_{42}$, $Pt_{52}Pd_{48}$, $Rh_{56}Ni_{44}$, and $Rh_{56}Pt_{44}$; however, CoFe, NiIr, NiMo, and NiNb were not alloyed because the reducing power of butyllithium used as a reducing agent was likely insufficient. To confirm whether the alloyed NPs have the same crystal structures as predicted by DFT, X-ray powder diffraction (XRD) patterns were investigated, as shown in Fig. 3. All of the samples showed alloyed diffraction patterns without signals from phases of the pure elements that compose each alloy. We also simulated XRD patterns of the alloys on the basis of the DFT crystal structures and found that the simulated patterns are matched well with the experimental patterns. This clearly reveals that we successfully synthesized alloyed NPs with the crystal structures predicted by DFT calculations for 9 bimetallic systems. The lattice parameters for the alloyed samples are presented in Table. 1 in Supporting Information. We further characterized the NP samples by a high-angle annular dark-field scanning transmission electron microscopy (HAADF-STEM) (Fig. 3 and Figs. S1-S9 in Supporting Information). The elemental color mapping and their overlay clearly indicate the homogeneous mixing of elements in each NP. The sizes of the alloyed NPs were in the range of 6.8±3.4 nm. For comparison, we also synthesized Pd NPs by the same method, and their sizes were 7.2±1.6 nm, which is in the range of the alloyed NPs (Fig. S10 in Supporting Information).

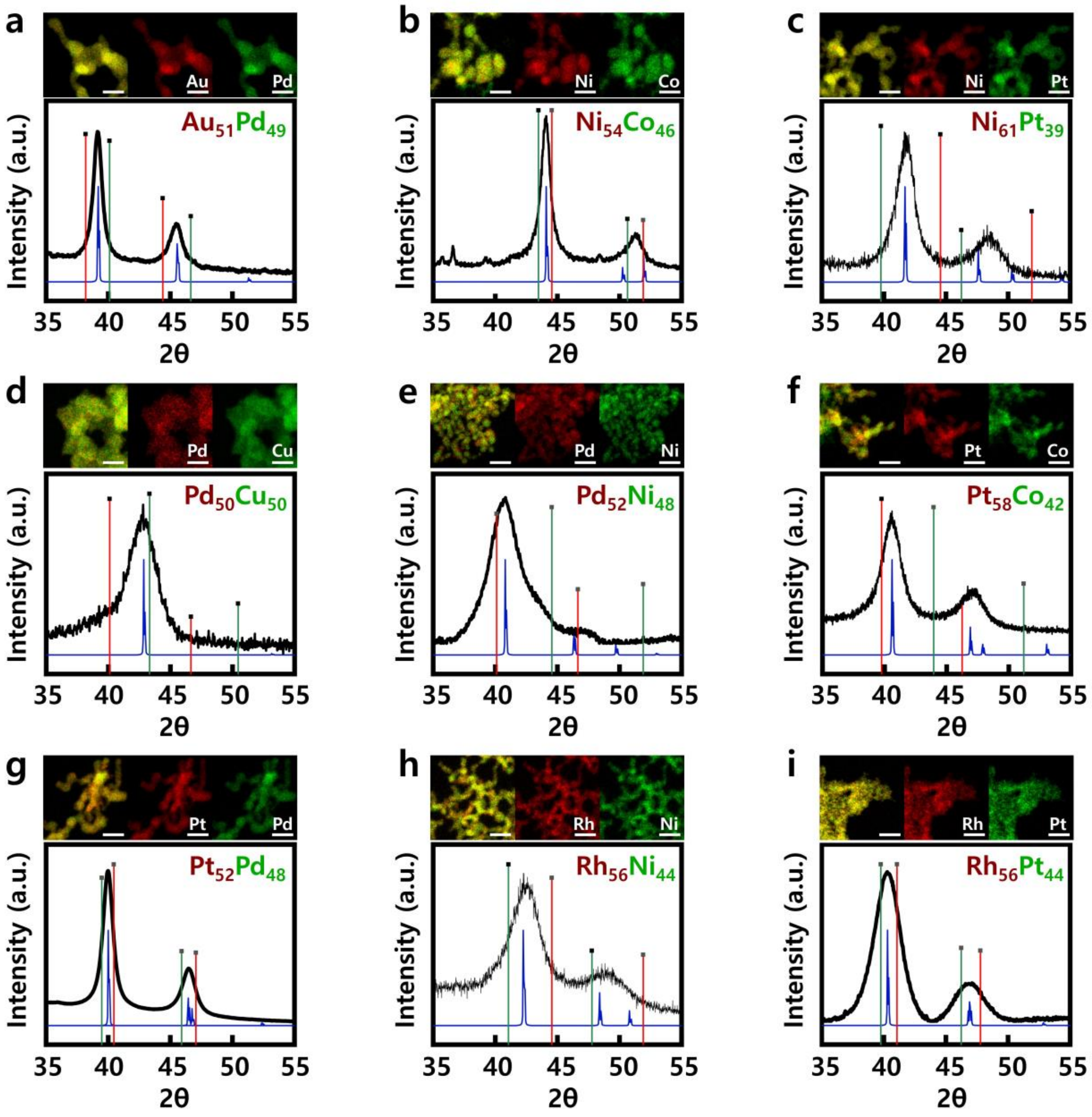

**Fig. 3 | Characterizations of the synthesized bimetallic NPs.** XRDs for **a.** $Au_{51}Pd_{49}$, **b.** $Ni_{54}Co_{46}$, **c.** $Ni_{61}Pt_{39}$, **d.** $Pd_{50}Cu_{50}$, **e.** $Pd_{52}Ni_{48}$, **f.** $Pt_{58}Co_{42}$, **g.** $Pt_{52}Pd_{48}$, **h.** $Rh_{56}Ni_{44}$, **i.** $Rh_{56}Pt_{44}$ NP samples. For comparison, the red and green lines in the XRD patterns correspond to the peaks of pure elements in each bimetallic NP. The blue peaks were simulated with the crystal structures of the bimetallic system predicted by DFT calculations. Above each XRD pattern, the element mapping images acquired by STEM-EDX are included, where the first image (the most left) is the overlay image of the maps for elements in each bimetallic NP. The scale bars correspond to 20 nm. The enlarged images of the STEM-EDX are also provided in Figs. S1-S9 of Supporting Information.

**Table 1 | Catalytic performances of 9 bimetallic NPs proposed by the DFT screening process in comparison to that of pure Pd NPs.**

| **Catalyst** | **$H_2O_2$ productivity** | | | **$H_2$ conversion (%)** | **$H_2O_2$ selectivity (%)** |
|---|---|---|---|---|---|
| | Mass-normalized (mmol/$g_{metal}$·h) | Cost-normalized (mmol/$\$_{metal}$·h) | TOF ($h^{-1}$) | | |
| $Au_{51}Pd_{49}$ (1.43)[a] | 505.2 | 7.1 | 77.1 | 11.6 | 70.1 |
| $Ni_{54}Co_{46}$ (1.71) | 10.2 | 31.5 | 0.6 | - | - |
| $Ni_{61}Pt_{39}$ (1.16) | 880.1 | 69.3 | 98.5 | 22.9 | 61.2 |
| $Pd_{50}Cu_{50}$ (1.51) | 49.4 | 1.3 | 4.2 | 1.6 | 50.8 |
| $Pd_{52}Ni_{48}$ (0.84) | 218.1 | 5.5 | 18.2 | 2.8 | >99.9[b] |
| $Pt_{58}Co_{42}$ (1.78) | 87.6 | 4.6 | 12.1 | 11.0 | 12.8 |
| $Pt_{52}Pd_{48}$ (1.16) | 1225.6 | 23.0 | 186.9 | 25.0 | 78.7 |
| $Rh_{56}Ni_{44}$ (0.98) | 16.8 | 0.1 | 1.4 | - | - |
| $Rh_{56}Pt_{44}$ (1.67) | 109.3 | 0.6 | 15.7 | 10.6 | 16.6 |
| **Pd** | 552.4 | 7.3 | 58.8 | 14.0 | 63.0 |

[a]The number in parentheses corresponds to $\Delta DOS_{2-1}$ for the alloy with a composition of 50:50 shown in Fig. 2.

[b]The high selectivity is mainly caused by the low $H_2$ conversion property.

To investigate whether the catalytic performance of the 9 proposed alloys is comparable to that of Pd, the alloyed catalysts were tested for direct $H_2O_2$ synthesis under mild conditions (20 °C, 1 atm) in comparison to pure Pd (Table 1). Along with the industrial importance of $H_2O_2$ as an eco-friendly oxidizing agent,[19,20] the $H_2O_2$ direct synthesis reaction can be performed under ambient temperature and pressure conditions, and Pd has been regarded as the archetypical catalyst for the chemical reaction although there is still room for performance enhancement.[21] In terms of $H_2O_2$ productivity representing the catalytic activity for $H_2O_2$ direct synthesis, three alloy systems, $Au_{51}Pd_{49}$ (505.2 $mmol_{H2O2}\cdot gmetal^{-1}\cdot h^{-1}$), $Ni_{61}Pt_{39}$ (880.1), and $Pt_{52}Pd_{48}$ (1225.6) exhibit performance comparable that of Pd (552.4). Additionally, $Pd_{52}Ni_{48}$ shows noticeable $H_2O_2$ productivity (218.1), although this productivity is lower than that of Pd. Here, it is notable that the 4 alloy systems have lower $\Delta DOS_{2-1}$ values than the alloy systems showing poor catalytic activities. Although the Rh-Ni system has a low $\Delta DOS_{2-1}$ value, the DOS patterns near $E_F$ are quite different from those of Pd (Fig. S11 in Supporting Information), which can explain why it shows poor catalytic activity. This result clearly demonstrates that a similarity in DOS patterns can be a good descriptor for the discovery of novel bimetallic catalysts, where no calculation relevant to the reaction pathways on the catalyst surfaces, which would require a high computational cost, has been performed.

Among the 4 alloy combinations showing the high $H_2O_2$ productivity, the Au-Pd,[13,14,22-24] Ni-Pd,[25] and Pd-Pt[26] systems were previously reported as promising candidates for $H_2O_2$ direct synthesis, validating our approach once again. In addition, our approach led to the discovery of a new catalyst, a Ni-Pt alloy, which has not yet been reported for $H_2O_2$ direct synthesis, although the system has been well known as an oxygen reduction reaction (ORR) catalyst in fuel cells.[27-29] As the price of Pd rapidly increases,[30] our Pd-free Ni-Pt alloy can serve as a more cost-effective option. To identify the cost-effectiveness of the Ni-Pt catalyst, we list the cost-normalized productivity (CNP; unit, $mmol_{H2O2}\cdot\$metal^{-1}\cdot h^{-1}$)[31] in Table 1, in which the metal price [unit, U.S. dollar ($)] instead of catalyst mass is used. In terms of this metric, $Ni_{61}Pt_{39}$ shows a CNP of 69.3, which is not only nearly 9.5 times higher than that of Pd (7.3) but also much higher than those of the other candidates ($Au_{51}Pd_{49}$: 7.1, $Pd_{52}Ni_{48}$: 5.5, and $Pt_{52}Pd_{48}$: 23.0). As another metric for catalytic activity, we also considered the turnover frequency (TOF) of the catalysts. Also in terms of this metric, $Ni_{61}Pt_{39}$ (98.5) shows a performance superior to that of Pd (58.8). Moreover, for efficient $H_2O_2$ direct synthesis, the $H_2$ conversion and $H_2O_2$ selectivity properties of the catalysts should be considered. It is also interesting that $Ni_{61}Pt_{39}$ shows the $H_2O_2$ selectivity similar to that of Pd, although $Ni_{61}Pt_{39}$

shows the higher $H_2$ conversion. These findings clearly demonstrate that the Ni-Pt bimetallic catalyst is very attractive for $H_2O_2$ direct synthesis.

When screening bimetallic catalysts by electronic structure calculations, only one composition of 50:50 for the bimetallic systems was considered because a DFT prediction to determine the optimized composition to provide the best catalytic performance would be challenging and time-consuming. Instead, the composition optimization proceeded with the help of experiments, likely providing a faster result than the DFT prediction. Therefore, to optimize the composition of the Ni-Pt bimetallic catalysts, we also explored the effects of the Ni/Pt ratio in $Ni_xPt_{100-x}$ NPs on their catalytic properties for $H_2O_2$ direct synthesis (Table S2 in Supporting Information), and no significant structural change in the $Ni_xPt_{100-x}$ NPs was not observed (Figs. S12 and S13 in Supporting Information). We found that the best catalytic performance was exhibited by the $Ni_{61}Pt_{39}$ composition in terms of $H_2O_2$ productivity, $H_2$ conversion, and $H_2O_2$ selectivity.

## Catalytic reaction mechanism

The Ni-Pt bimetallic catalysts were proposed based only on electronic structure calculations without any calculations relevant to the reaction pathways. Thus, to justify the experimental catalytic activity shown in Table 1, we investigated the reaction mechanisms on the catalyst surface using DFT calculations. We considered the close-packed (111) surface of $Ni_{50}Pt_{50}$ slab in the $L1_0$, where the alloy composition was considered for simplicity of the calculations although the best catalytic performances were experimentally observed for the $Ni_{61}Pt_{39}$ composition. The Langmuir-Hinshelwood mechanism[32-34] involving adsorption of both $H_2$ and $O_2$ over a catalyst surface was first explored. Our DFT calculations revealed that dissociative adsorption of $H_2$ occurs very favorably over $Ni_{50}Pt_{50}(111)$ surfaces (Fig. S14 in Supporting Information), which is similar to Pd(111).[32,33] This implies that the hydrogenation of $O_2$ over Pd(111) and $Ni_{50}Pt_{50}(111)$ determines the catalytic activity for $H_2O_2$ synthesis. In Fig. 4a, the overall $H_2O_2$ synthesis involving two sequential hydrogenations of $O_2$ (I → II and II → III) is exothermic by 0.42 eV on $Ni_{50}Pt_{50}(111)$, which is more favorable than the value of 0.07 eV for Pd(111). Moreover, while the rate-determining step on Pd(111) is the first hydrogenation (I → II) to form OOH* (* denotes a surface site) with an energy barrier of 0.78 eV, the rate-determining step is observed at the desorption step of the generated $H_2O_2$*, with an energy barrier of 0.69 eV (III → V), on $Ni_{50}Pt_{50}(111)$. These results indicate that the higher $H_2O_2$ productivity is possible on $Ni_{61}Pt_{39}$ NPs than on Pd NPs, as observed in Table 1. On the other hand, Wilson and Flaherty proposed another mechanism for $H_2O_2$ formation on Pd NPs called a heterolytic reaction mechanism,[35] similar to the two-electron oxygen reduction reaction (ORR), in which an electron is generated during the oxidation process of $H_2$ on the NP surfaces. We also investigated the mechanism on $Ni_{50}Pt_{50}$(111) and Pd(111) by DFT calculations where the computational hydrogen electrode method was used along with consideration of implicit water solvents (Fig. 4b). On both Pd(111) and $Ni_{50}Pt_{50}$(111), the ORR processes for $H_2O_2$ formation are thermodynamically favorable. In particular, although the first protonation is very similar (-1.03 eV for Pd and -1.00 eV for $Ni_{50}Pt_{50}$), the second protonation is slightly more favorable on $Ni_{50}Pt_{50}$(111) (-1.40 eV vs. -1.58 eV). This mechanism also readily reveals more preferential $H_2O_2$ production on the Ni-Pt surfaces than on the Pd surfaces.

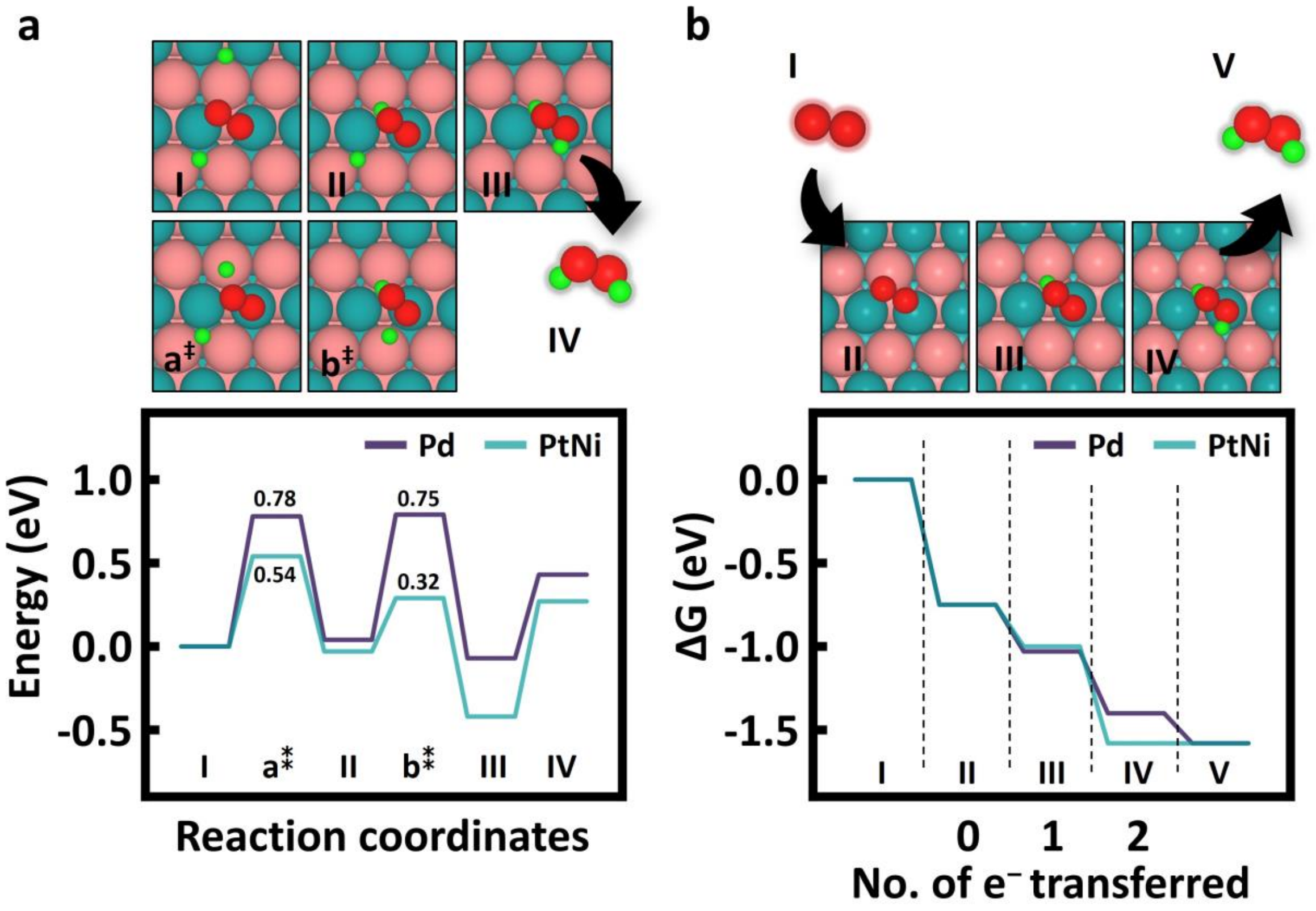

**Fig. 4 | $H_2O_2$ direct synthesis mechanisms of $Ni_{50}Pt_{50}$(111) and Pd(111) from DFT calculations. a.** The Langmuir-Hinshelwood mechanism. **b.** The electron-proton transfer mechanism. The cyan lines are the energy diagram for $Ni_{50}Pt_{50}$(111), and the magenta lines are for Pd(111). The color code for the atoms is as follows: teal = Ni, coral = Pt, red = O, and green = H.

## Conclusion

This work demonstrates an efficient computational-experimental screening protocol for the development of bimetallic catalysts that replace conventional Pd catalysts for $H_2O_2$ direct synthesis. Using this protocol, a novel Pd-free bimetallic catalyst, $Ni_{61}Pt_{39}$, was successfully developed. Here, we note that Ni and Pt have not usually been considered in the catalysis field for $H_2O_2$ direct synthesis because they prefer to dissociate $O_2$ molecules, which prevents $H_2O_2$ formation, although the Ni-Pd,[25] Pt-Pd,[26] and Pt-Ag[36] bimetallic catalysts have been reported. The successful story of our protocol basically results from the validity of our descriptor, the similarity of electronic DOS patterns, for catalyst screening. In this regard, this protocol will immediately provide new opportunities for catalyst discovery for the replacement or reduction in use of the platinum-group metals and then readily apply them into other chemical reactions.

## Methods

**Computational details for screening bimetallic catalysts.** We generated 10 ordered phases for each bimetallic alloy with a 1:1 composition. Specifically, the crystal structure prototypes were obtained by NaCl (B1), CsCl (B2), zinc blende (B3), wurtzite (B4), TiCu (B11), CdAu (B19), FeB (B27), CrB (B33), AuCu ($L1_0$), and CuPt ($L1_1$) compounds from the Material Project.[37] Here, the initial lattice parameters were replaced by values commensurate with covalent radii of two elements in the target compounds. By means of the DFT optimization process, we collected the most stable bulk phases of 4,350 binary alloys. Furthermore, the formation energies ($\Delta E_f$) of all the bulk phases were calculated as follows: $\Delta E_f = E[AB] - E[A] - E[B]$, where E[AB], E[A], and E[B] are the total energies of the bimetallic AB compound, A crystal, and B crystal in unit cell. For thermodynamically favorable alloys, a close-packed surface of each crystal structure was used for the calculation of the DOS pattern of each alloy. In principle, the close-packed surface of $L1_0$ is the (111) facet, that of B2 is the (110) facet, and those of other crystal structures are the (001) facet.[38]

To investigate thermodynamic stabilities in bulk phases and the DOS patterns of surface atoms, spin-polarized DFT calculations were performed using the Vienna Ab Initio Simulation Package (VASP)[39] with projector-augmented-wave (PAW) pseudopotentials[40] and the revised Perdew-Burke-Ernzerhof (RPBE) exchange-correlation functional.[41] Then, we used a plane-wave kinetic energy cutoff of 400 eV and Monkhorst-Pack k-point meshes of 8×8×8 for the bulk systems and 4×4×1 for the slab systems. The bulk of the 1×1×1 unit cell and the slab of the 2×2 unit cell with 4 layers were used. In the slab systems, a vacuum spacing of 15 Å was used to avoid interactions between slabs. All relaxations were performed until the energy change was less than $1 \times 10^{-6}$ eV/cell and the forces on the individual atoms were less than 0.01 eV $Å^{-1}$.

**$H_2O_2$ formation mechanism.** The energy diagrams for $H_2O_2$ production on the $Ni_{50}Pt_{50}(111)$ surface and Pd(111) surfaces were also calculated by the DFT method. The simulations were performed using the VASP[39] using the PAW[40] pseudopotentials to describe the potential from the ionic core. For the exchange and correlation terms, we employed the RPBE functional[41] with Grimme's D3 dispersion correction.[42] An energy cut-off of 400 eV and Monkhorst-Pack k-point meshes of 3×3×1 for the slab calculations were used. A large vacuum spacing of 15 Å was used to prevent interslab interactions. The top two layers and the reactant molecules were optimized until the energy change was less than $1 \times 10^{-4}$ eV/cell and the force on each atom was less than 0.02 eV $Å^{-1}$. In the case of the Langmuir-Hinshelwood mechanism, the climbing image nudged elastic band (CI-NEB)[43] method was employed to investigate the transition states for the main reaction. For the electron-proton transfer mechanism, the computational hydrogen electrode method[44] with $U = 0$ V in addition to the water solvation energy calculated with the implicit solvent model implemented within the $VASP_{sol}$ package[45] was used to construct the energy diagram on the following reactions:

$$O_{2(g)} \rightarrow {}^*O_2 \quad (1)$$

$$^*O_2 + H^+ + e^- \rightarrow {}^*OOH \quad (2)$$

$$^*OOH + H^+ + e^- \rightarrow {}^*H_2O_2 \quad (3)$$

$$^*H_2O_2 \rightarrow H_2O_{2(g)} \quad (4)$$

**Chemicals.** Palladium(II) acetate [$Pd(CH_3COO)_2$, ≥99.9%], copper(II) acetate [$Cu(CH_3COO)_2$, 99.99%], nickel(II) acetylacetonate [$Ni(acac)_2$, 95%], rhodium(III) nitrate hydrate [$Rh(NO_3)_3 \cdot XH_2O$, ~36%], gold(III) chloride trihydrate [$HAuCl_4 \cdot 3H_2O$, 99.9%], chloroplatinic acid hydrate [$H_2PtCl_6 \cdot XH_2O$, ~38%], cobalt acetate tetrahydrate [$Co(CH_3COO) \cdot 4H_2O$, 99.999%], dioctyl ether, oleylamine, and n-butyllithium solution [2.0 M in cyclohexane] were purchased from Sigma-Aldrich. Ethanol was analytical grade and was used without further purification.

**Catalyst synthesis.** All alloy metal catalysts and Pd catalysts were synthesized using the butyllithium reduction method. To synthesize the bimetallic XY catalysts, 0.017 mmol of X precursor and Y precursor (details are described in Table 3 of Supporting Information) were dissolved in 10 mL of dioctyl ether with 2 mL of oleylamine at 50 °C. The solution was then injected into a butyllithium solution containing 15 mL of dioctyl ether and 1.2 mL of 2.0 M butyllithium in cyclohexane at 50 °C. The colloids were stirred (500 rpm) for 20 min and then heated to 120 °C for 1.5 h in an Ar atmosphere. The reaction mixture was further heated to 260 °C for 1 h. The mixture was cooled to room temperature, and 1.25 mL of trioctylphosphine was injected to protect the colloids. The resulting NPs were washed with ethanol 3 times.

**Catalyst characterization.** Transmission electron microscopy (TEM) images and energy–dispersive X–ray spectra (EDS) were acquired using a transmission electron microscope (FEI Talos F200X) equipped with scanning transmission electron microscopy–energy dispersive X–ray spectroscopy (Bruker Super–X EDS system). The crystal structure was examined by X–ray diffraction (Rigaku Dmax 2500) with Cu Kα radiation ($\lambda$ = 1.5406 Å). The samples for the TEM images and X-ray diffraction patterns were purified by centrifugation three times to remove the surfactants and/or excess reactants. Then, the NPs were dispersed in ethanol. The resulting solution was dropped on a copper grid coated with an amorphous carbon film for the TEM images and a soda-lime-silica glass for the X-ray diffraction pattern. The actual compositions of each alloy catalyst were measured via inductively coupled plasma atomic emission spectrometry (ICP-OES) using iCAP6500 Duo (Thermo).

**Catalyst performance measurements.** The direct synthesis of $H_2O_2$ was performed in a double-jacket glass reactor under stirring at 1200 rpm. The reaction medium (150 ml) was composed of ethanol and water (ethanol:water volume ratio = 1:4) with 0.9 mM sodium bromide (NaBr) and 0.02 M phosphoric acid ($H_3PO_4$). The catalyst weight used was 1 mg. The volumetric flow rate of the reactant gas stream was 22 ml/min ($H_2$:$O_2$ volume ratio = 1:10). The reaction was performed at 293 K and 1 atm for 1 h. After the reaction, the concentration of $H_2O_2$ was calculated using iodometric titration. The concentration of $H_2$ was calculated using gas chromatography (Younglin, 6500GC) equipped with a Molecular Sieve 5A (Supelco, 60/80 mesh, 6 ft x 1/8 in, 2.1 mm) packed column and a thermal conductive detector (TCD). The catalytic properties ($H_2O_2$ productivity, $H_2$ conversion, and $H_2O_2$ selectivity) shown in Table 1 were calculated with the following equations:

$$H_2 \text{ conversion } (\%) = \frac{\text{moles of reacted } H_2}{\text{moles of supplied } H_2} \times 100$$

$$H_2O_2 \text{ selectivity } (\%) = \frac{\text{moles of formed } H_2O_2}{\text{moles of reacted } H_2} \times 100$$

$$H_2O_2 \text{ productivity } (\text{mmol/g}_{\text{metal}}\cdot\text{h}) = \frac{\text{mmoles of formed } H_2O_2}{\text{weight of metal (g)} \times \text{reaction time (h)}}$$

$$\text{TOF } (\text{h}^{-1}) = \frac{\text{mmoles of formed } H_2O_2}{\text{mmols of metal } \times \text{reaction time (h)}}$$

## ASSOCIATED CONTENT

### Supporting Information

The Supporting Information is available free of charge at http://pubs.acs.org

Characterization results (crystal structures, lattice parameters, HAADF-STEM and STEM-EDX) and weights of metal precursors used for the synthesis of the 9 bimetallic ($Au_{51}Pd_{49}$, $Ni_{54}Co_{46}$, $Ni_{61}Pt_{39}$, $Pd_{50}Cu_{50}$, $Pd_{52}Ni_{48}$, $Pt_{58}Co_{42}$, $Pt_{52}Pd_{48}$, $Rh_{56}Ni_{44}$, and $Rh_{56}Pt_{44}$) NPs; electronic DOS patterns of the 9 bimetallic surfaces; characterization results (XRDs, HAADF-STEM, and STEM-EDX) of $Ni_xPt_{100-x}$ NPs; DFT calculations of $H_2$ dissociation over $Ni_{50}Pt_{50}(111)$.

## AUTHOR INFORMATION

### Corresponding Authors

**Seung Yong Lee** – Materials Achitecturing Research Center, Korea Institute of Science and Technology, Seoul 02792, Republic of Korea; Email: patra@kist.re.kr

**Sang Soo Han** – Computational Science Research Center, Korea Institute of Science and Technology, Seoul 02792, Republic of Korea; Email: sangsoo@kist.re.kr

**Authors**

**Byung Chul Yeo** – Computational Science Research Center, Korea Institute of Science and Technology, Seoul 02792, Republic of Korea; Department of Energy Resources Engineering, Pukyong National University, Busan 48513, Republic of Korea

**Hyunji Nam** – Materials Achitecturing Research Center, Korea Institute of Science and Technology, Seoul 02792, Republic of Korea; Department of Chemical and Biological Engineering, Korea University, Seoul 02841, Republic of Korea

**Hyobin Nam** – Materials Achitecturing Research Center, Korea Institute of Science and Technology, Seoul 02792, Republic of Korea; Department of Nanomaterials Science and Engineering, Korea University of Science and Technology, Daejeon 34113, Republic of Korea

**Min-Cheol Kim** – Computational Science Research Center, Korea Institute of Science and Technology, Seoul 02792, Republic of Korea

**Hong Woo Lee** – Computational Science Research Center, Korea Institute of Science and Technology, Seoul 02792, Republic of Korea

**Donghun Kim** – Computational Science Research Center, Korea Institute of Science and Technology, Seoul 02792, Republic of Korea

**Kwan-Young Lee** – Department of Chemical and Biological Engineering, Korea University, Seoul 02841, Republic of Korea

**Author Contributions**

[†] Byung Chul Yeo and Hyunji Nam: These authors contributed equally to this work.

**Notes**

The authors declare no competing financial interest

**ACKNOWLEDGEMENTS**

This work was supported by Creative Materials Discovery Program through the National Research Foundation of Korea (NRF-2016M3D1A1021141). We acknowledge the financial supports of the Korea Institute of Science and Technology (Grant No. 2E30460).

## Table of Contents Graphic

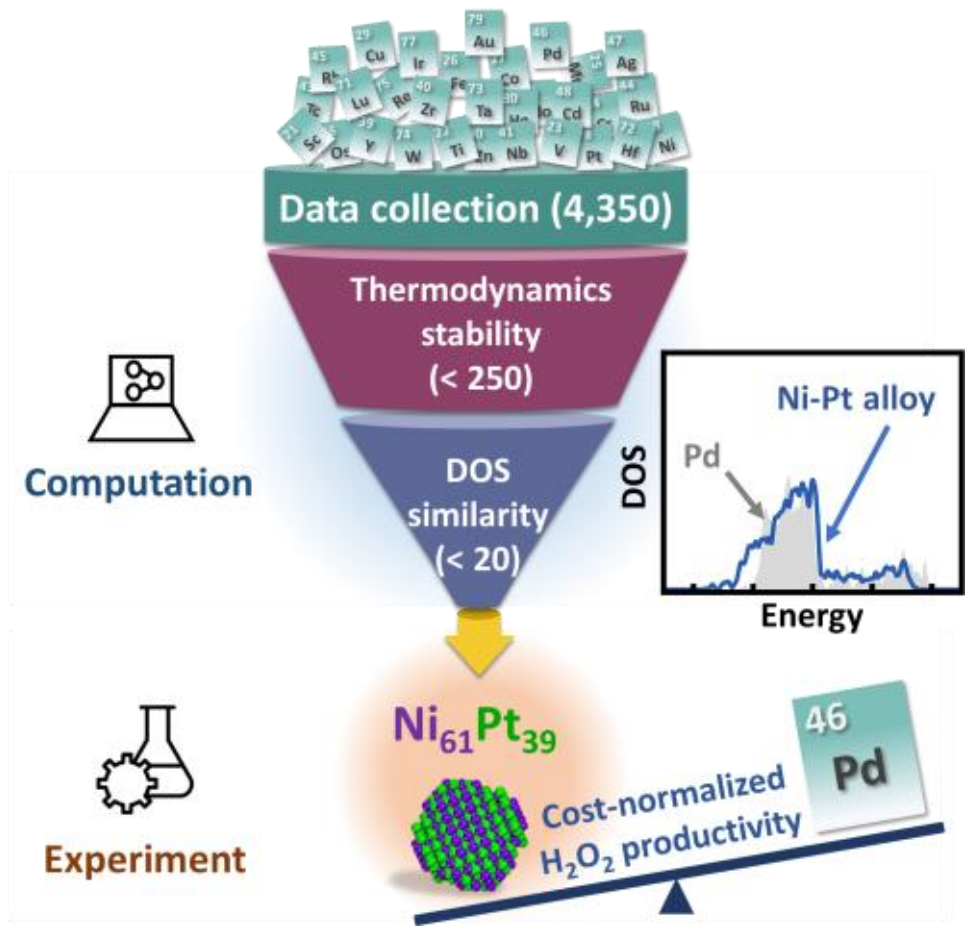

Supporting Information for

# High-Throughput Computational-Experimental Screening Protocol for the Discovery of Bimetallic Catalysts

Byung Chul Yeo,[1,2,†] Hyunji Nam,[3,4,†] Hyobin Nam,[3,5] Min-Cheol Kim,[1] Hong Woo Lee,[1,4] Donghun Kim,[1] Kwan-Young Lee,[4] Seung Yong Lee,[3,5,*] and Sang Soo Han[1,*]

[1]Computational Science Research Center, Korea Institute of Science and Technology, Seoul 02792, Republic of Korea
[2]Department of Energy Resources Engineering, Pukyong National University, Busan 48513, Republic of Korea
[3]Materials Achitecturing Research Center, Korea Institute of Science and Technology, Seoul 02792, Republic of Korea
[4]Department of Chemical and Biological Engineering, Korea University, Seoul 02841, Republic of Korea
[5]Department of Nanomaterials Science and Engineering, Korea University of Science and Technology, Daejeon 34113, Republic of Korea

*Correspondence to: sangsoo@kist.re.kr (S.S.H); patra@kist.re.kr (S.Y.L)
[†] These authors contributed equally to this work.

**Table S1. Crystal structures and lattice parameters of the 9 synthesized bimetallic NPs. For comparison, the Pd NP information is included**.

| Catalyst | Crystal structure | Lattice constant (Å) |
|---|---|---|
| $Au_{51}Pd_{49}$ | $L1_0$ | 3.98 |
| $Ni_{54}Co_{46}$ | $L1_0$ | 3.54 |
| $Ni_{61}Pt_{39}$ | $L1_0$ | 3.75 |
| $Pd_{50}Cu_{50}$ | B2 | 2.98 |
| $Pd_{52}Ni_{48}$ | $L1_0$ | 3.80 |
| $Pt_{58}Co_{42}$ | $L1_0$ | 3.85 |
| $Pt_{52}Pd_{48}$ | $L1_0$ | 3.90 |
| $Rh_{56}Ni_{44}$ | $L1_0$ | 3.70 |
| $Rh_{56}Pt_{44}$ | $L1_0$ | 3.87 |
| **Pd** | FCC | 3.87 |

**Table S2. Catalytic effects of Ni/Pt ratios in $Ni_xPt_{100-x}$ NPs for $H_2O_2$ direct synthesis.**

| | $H_2O_2$ productivity (mmol/g·h) | $H_2$ conversion (%) | $H_2O_2$ selectivity (%) |
|---|---|---|---|
| $Ni_{94}Pt_6$ | 338.7 | 13.8 | 44.5 |
| $Ni_{79}Pt_{21}$ | 502.9 | 15.4 | 52.3 |
| $Ni_{61}Pt_{39}$ | 880.1 | 22.9 | 61.5 |
| $Ni_{38}Pt_{62}$ | 473.5 | 19.5 | 38.8 |
| $Ni_9Pt_{91}$ | 673.1 | 21.5 | 50.1 |

**Table S3. Weight of metal precursors used for the synthesis of the 9 bimetallic and Pd NPs.**

| | **Precursor of X / weight (mg)** | **precursor of Y / weight (mg)** |
|---|---|---|
| **$Au_{51}Pd_{49}$** | $HAuCl_4 \cdot 3H_2O$ / **67** | $Pd(CH_3COO)_2$ / **38** |
| **$Ni_{54}Co_{46}$** | $Ni(acac)_2$ / **42** | $Co(CH_3COO) \cdot 4H_2O$ / **42** |
| **$Ni_{61}Pt_{39}$** | $Ni(acac)_2$ / **42** | $H_2PtCl_6$ / **70** |
| **$Pd_{50}Cu_{50}$** | $Pd(CH_3COO)$ / **38** | $Cu(CH_3COO)_2$ / **31** |
| **$Pd_{52}Ni_{48}$** | $Pd(CH_3COO)$ / **38** | $Ni(acac)_2$ / **42** |
| **$Pt_{58}Co_{42}$** | $H_2PtCl_6$ / **70** | $Co(CH_3COO) \cdot 4H_2O$ / **42** |
| **$Pt_{52}Pd_{48}$** | $H_2PtCl_6$ / **70** | $Pd(CH_3COO)_2$ / **38** |
| **$Rh_{56}Ni_{44}$** | $Rh(NO_3)_3$ / **49** | $Ni(acac)_2$ / **42** |
| **$Rh_{56}Pt_{44}$** | $Rh(NO_3)_3$ / **49** | $H_2PtCl_6$ / **70** |
| **Pd** | $Pd(CH_3COO)$ / **76** | |

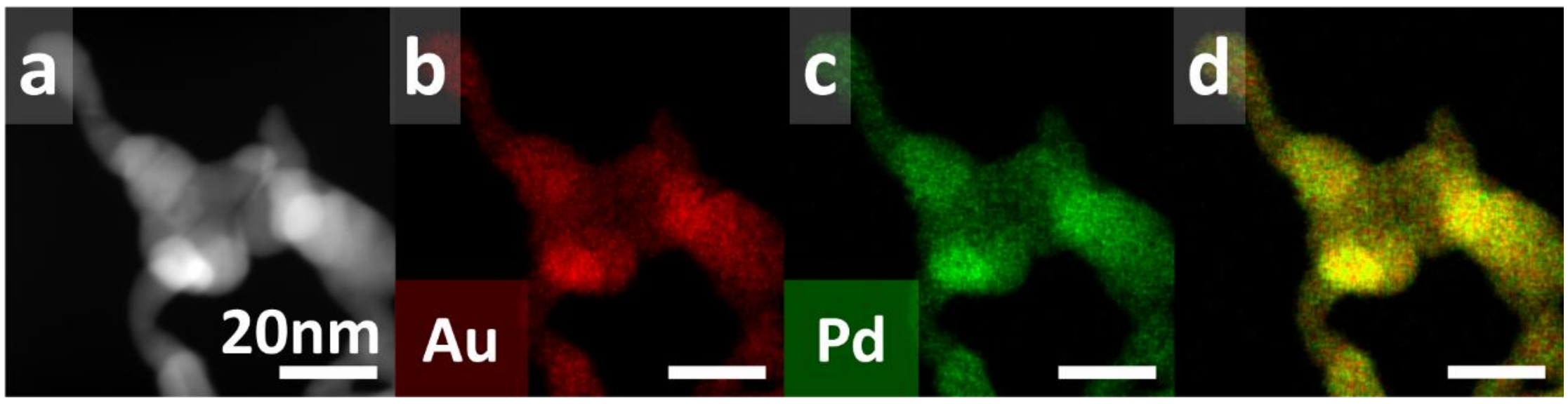

**Figure S1. Characterization of $Au_{51}Pd_{49}$. a**. HAADF-STEM image, **b**. Au-L STEM-EDX map, **c**. Pd-L STEM-EDX map, and **d**. the reconstructed overlay image of the maps shown in panels b and c (red, Au; green, Pd). The scale bars correspond to 20 nm.

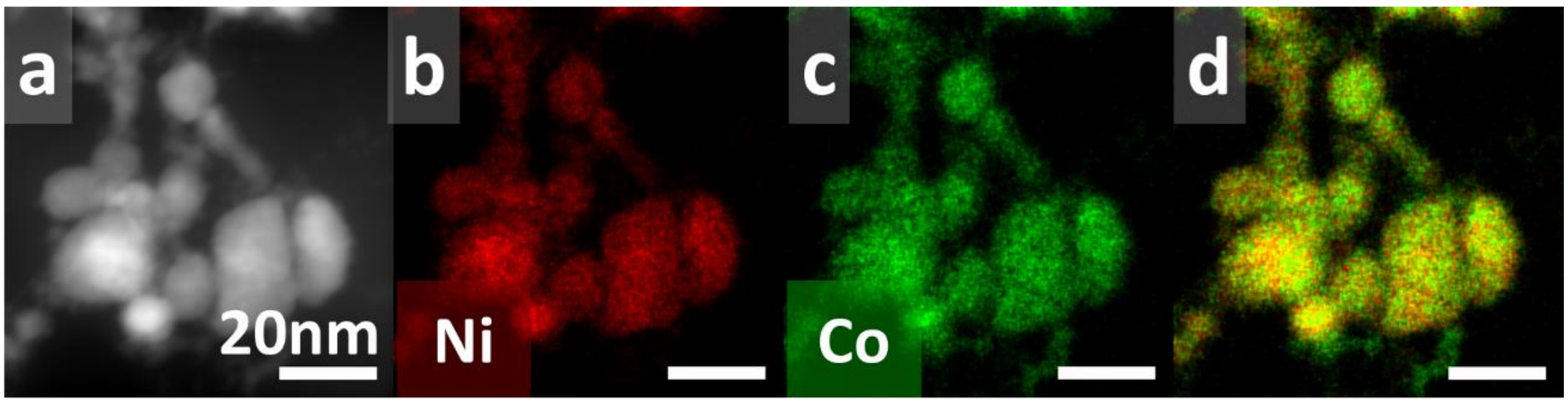

**Figure S2. Characterization of $Ni_{54}Co_{46}$. a**. HAADF-STEM image, **b**. Ni-K STEM-EDX map, **c**. Co-K STEM-EDX map, and **d**. the reconstructed overlay image of the maps shown in panels b and c (red, Ni; green, Co). The scale bars correspond to 20 nm.

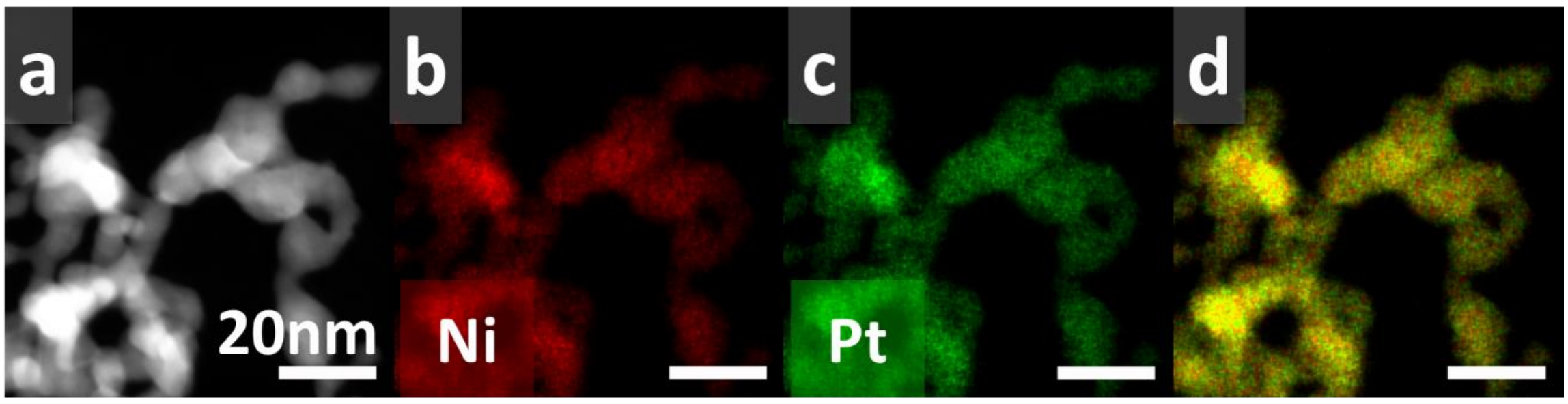

**Figure S3. Characterization of $Ni_{61}Pt_{39}$. a**. HAADF-STEM image, **b**. Ni-K STEM-EDX map, **c**. Pt-L STEM-EDX map, and **d**. the reconstructed overlay image of the maps shown in panels b and c (red, Ni; green, Pt). The scale bars correspond to 20 nm.

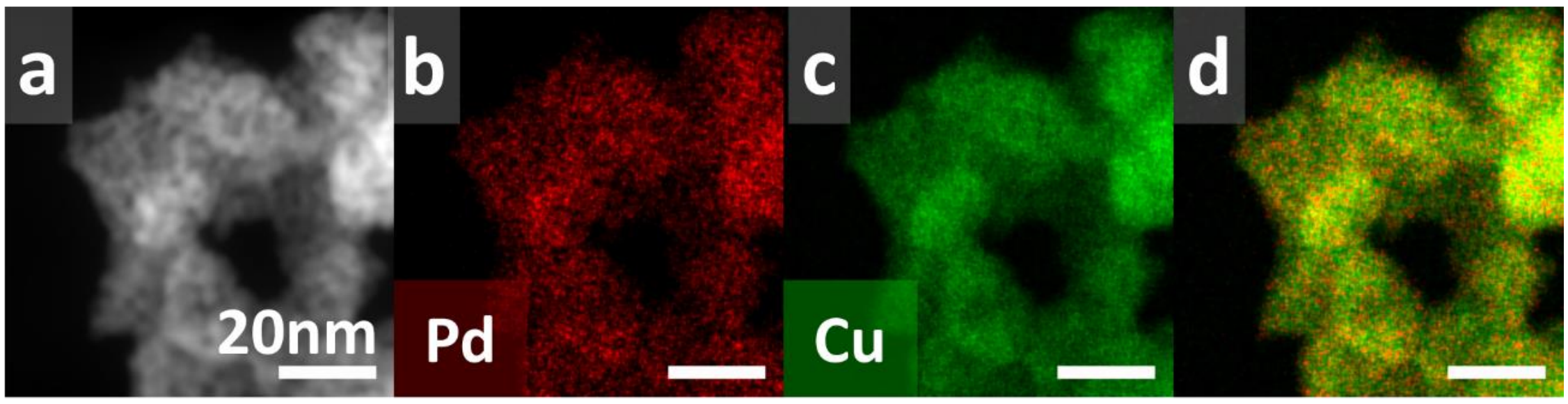

**Figure S4. Characterization of $Pd_{50}Cu_{50}$.** **a**. HAADF-STEM image, **b**. Pd-L STEM-EDX map, **c**. Cu-K STEM-EDX map, and **d**. the reconstructed overlay image of the maps shown in panels b and c (red, Pd; green, Cu). The scale bars correspond to 20 nm.

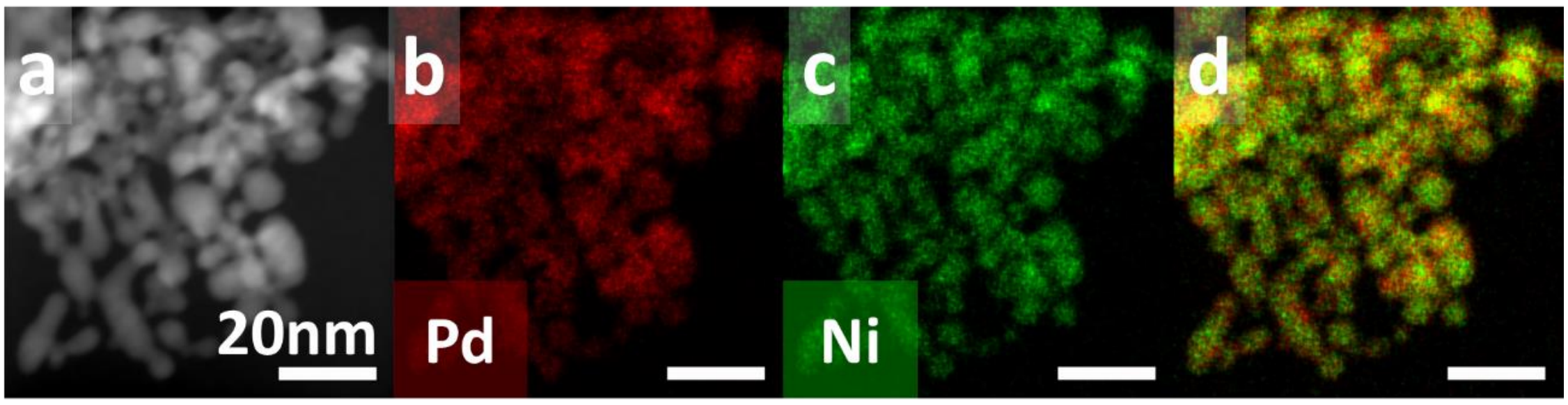

**Figure S5. Characterization of $Pd_{52}Ni_{48}$.** **a**. HAADF-STEM image, **b**. Pd-L STEM-EDX map, **c**. Ni-K STEM-EDX map, and **d**. the reconstructed overlay image of the maps shown in panels b and c (red, Pd; green, Ni). The scale bars correspond to 20 nm.

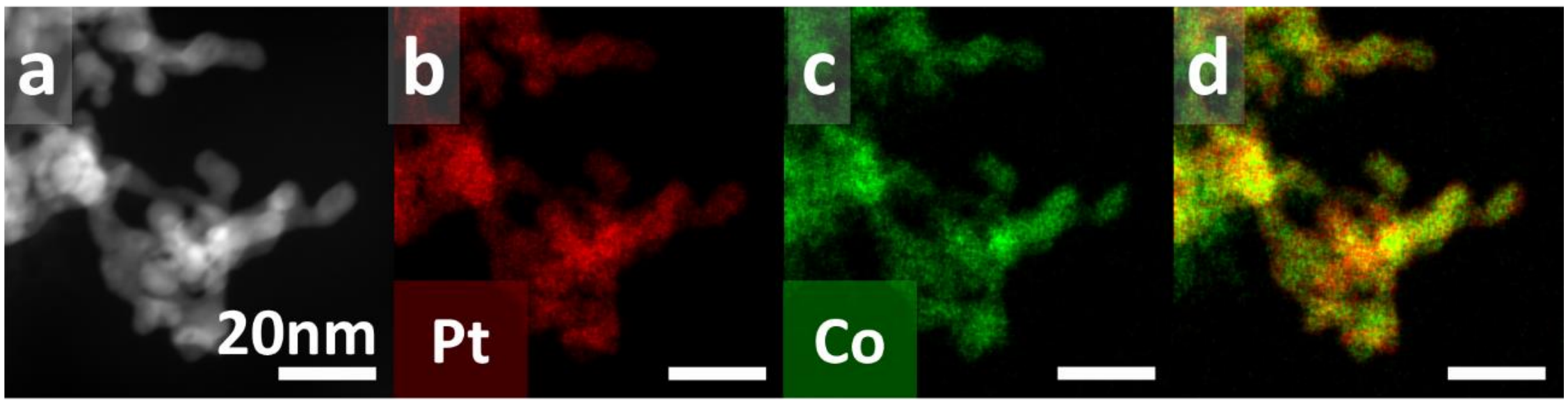

**Figure S6. Characterization of $Pt_{58}Co_{42}$.** **a**. HAADF-STEM image, **b**. Pt-L STEM-EDX map, **c**. Co-K STEM-EDX map, and **d**. the reconstructed overlay image of the maps shown in panels b and c (red, Pt; green, Co). The scale bars correspond to 20 nm.

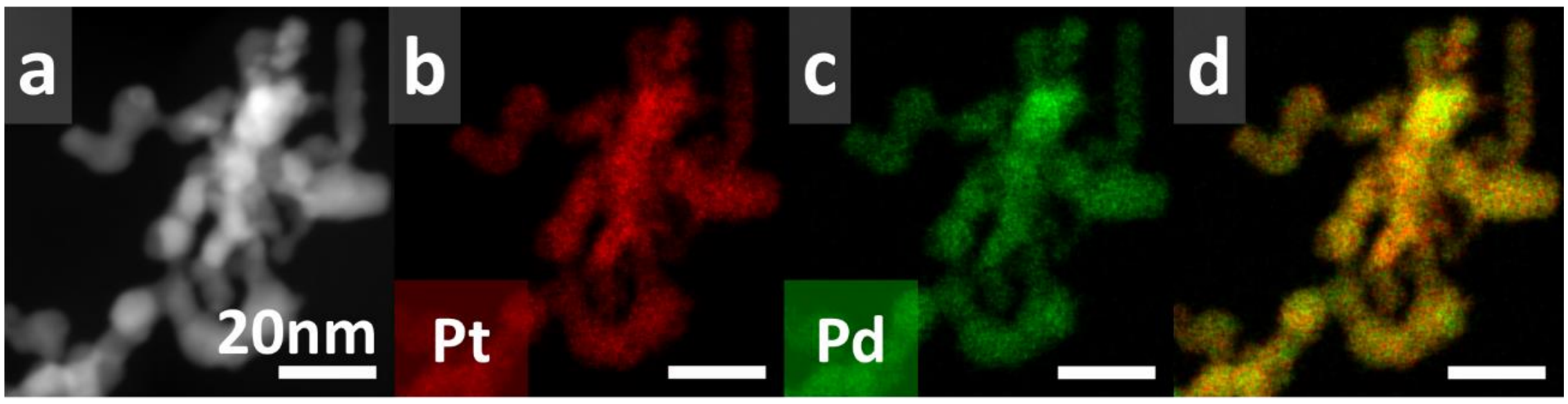

**Figure S7. Characterization of $Pt_{52}Pd_{48}$.** **a**. HAADF-STEM image, **b**. Pt-L STEM-EDX map, **c**. Pd-L STEM-EDX map, and **d**. the reconstructed overlay image of the maps shown in panels b and c (red, Pt; green, Pd). The scale bars correspond to 20 nm.

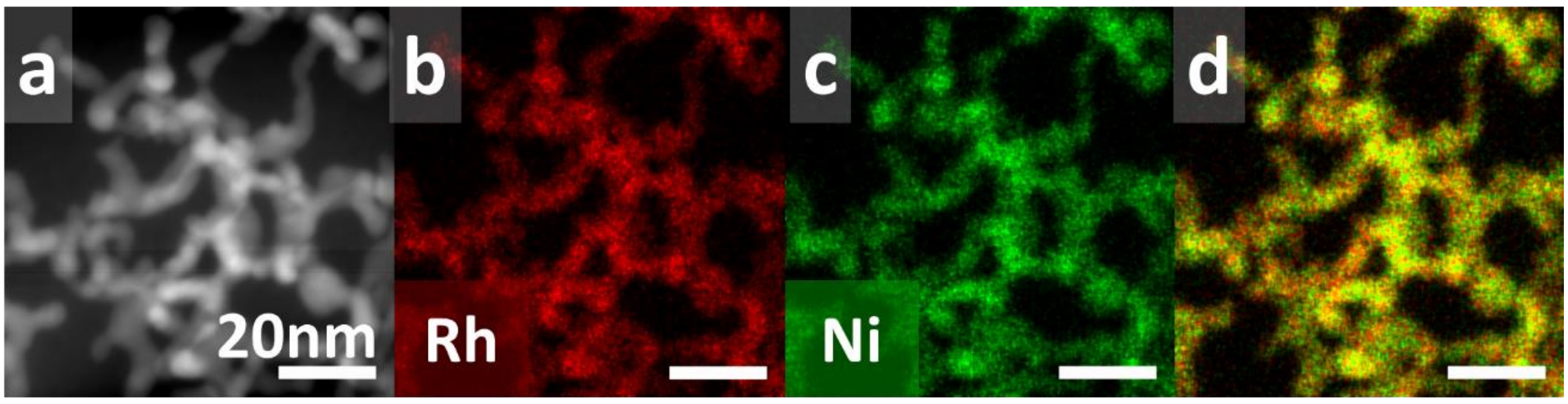

**Figure S8. Characterization of $Rh_{56}Ni_{44}$.** **a**. HAADF-STEM image, **b**. Rh-L STEM-EDX map, **c**. Ni-K STEM-EDX map, and **d**. the reconstructed overlay image of the maps shown in panels b and c (red, Rh; green, Ni). The scale bars correspond to 20 nm.

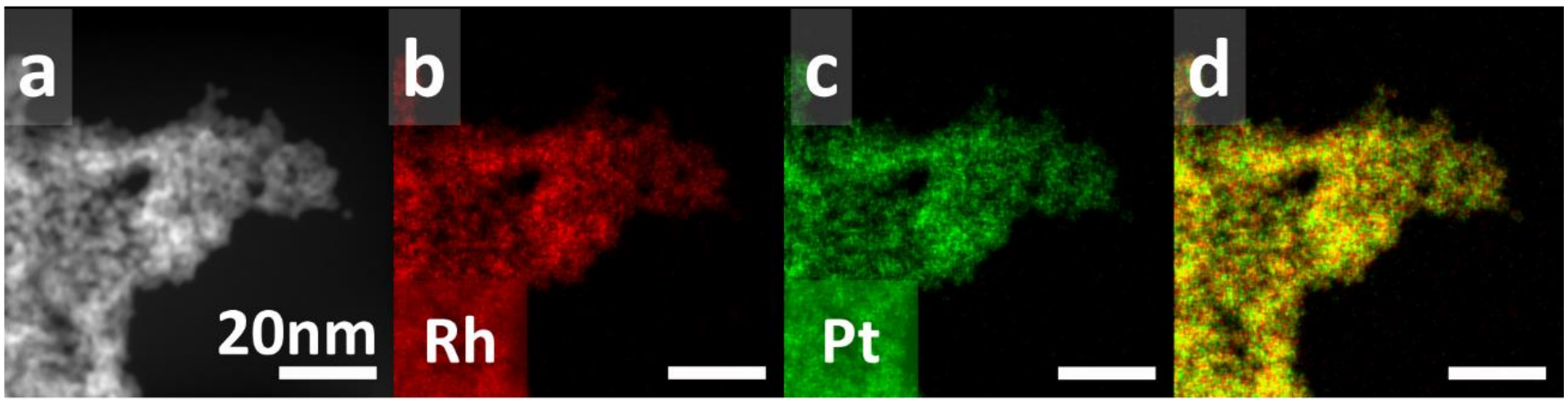

**Figure S9. Characterization of $Rh_{56}Pt_{44}$.** **a**. HAADF-STEM image, **b**. Rh-L STEM-EDX map, **c**. Pt-L STEM-EDX map, and **d**. the reconstructed overlay image of the maps shown in panels b and c (red, Rh; green, Pt). The scale bars correspond to 20 nm.

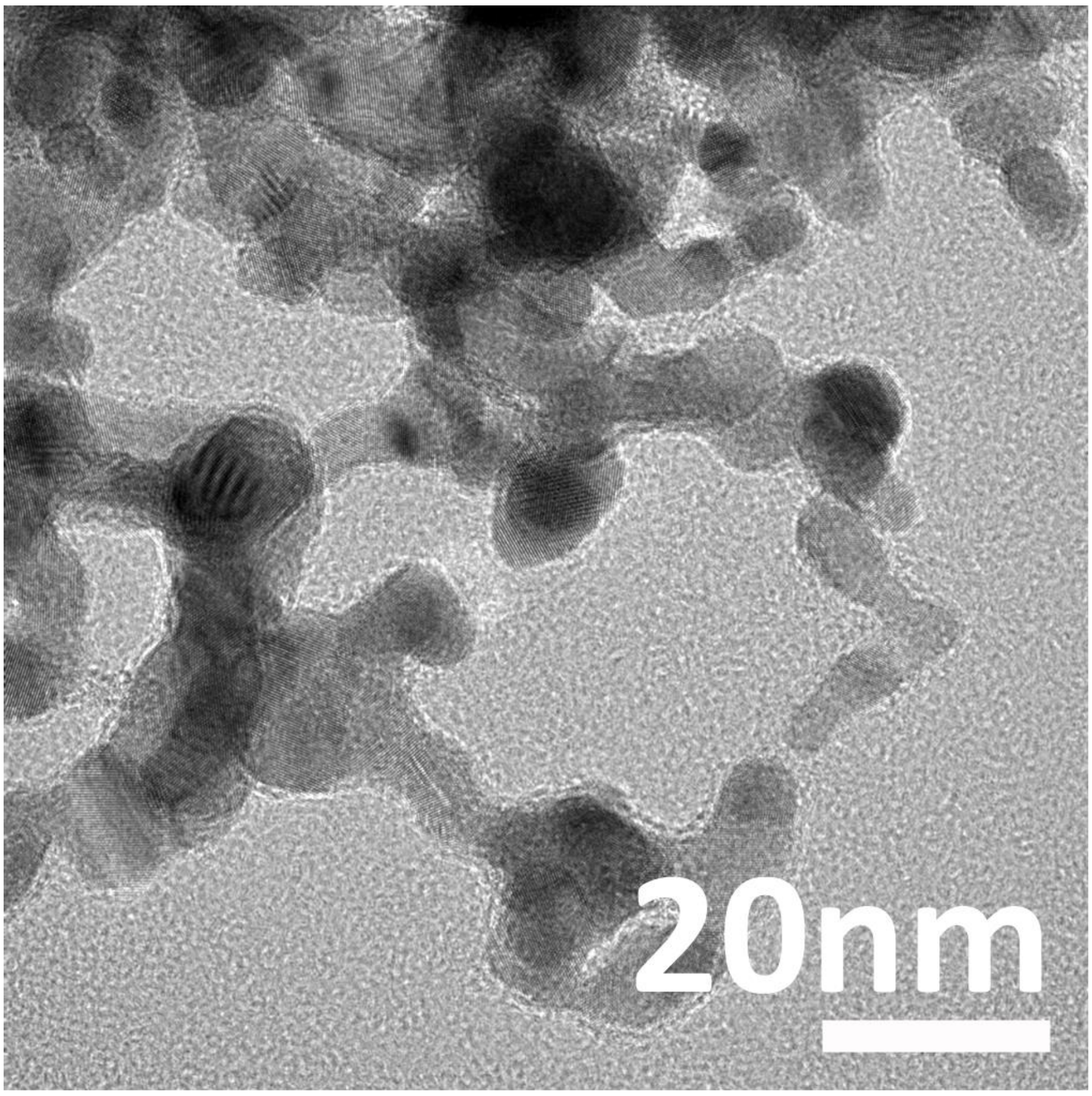

**Figure S10. HAADF-STEM image of Pd.** The Pd NPs synthesized by a butyllithium reduction method in this work show an average size of 7.2 nm.

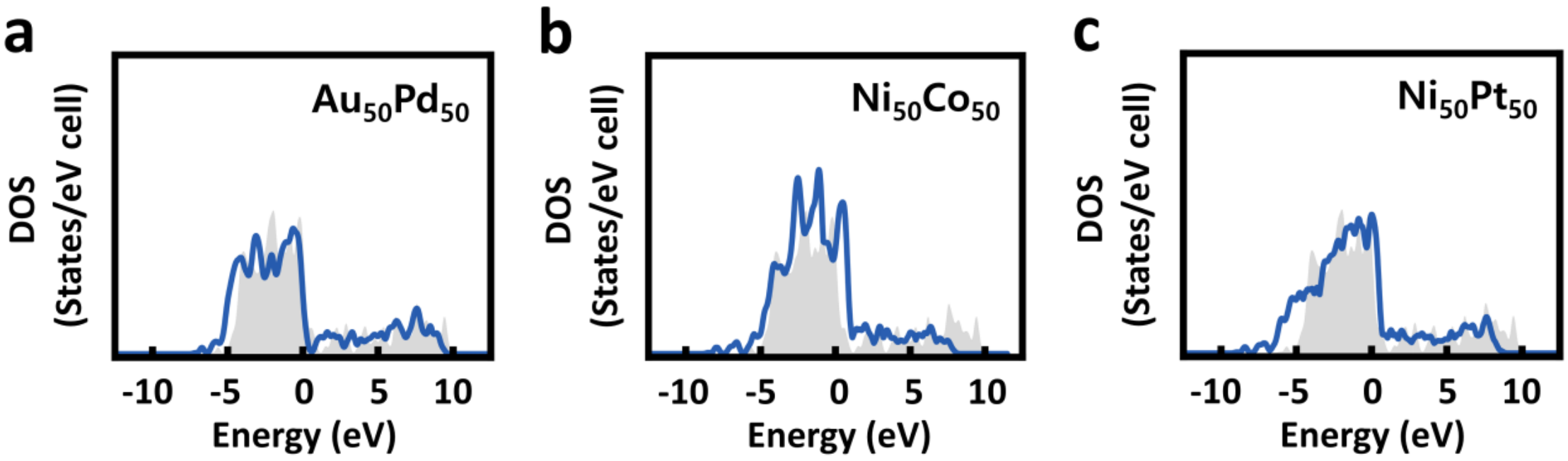

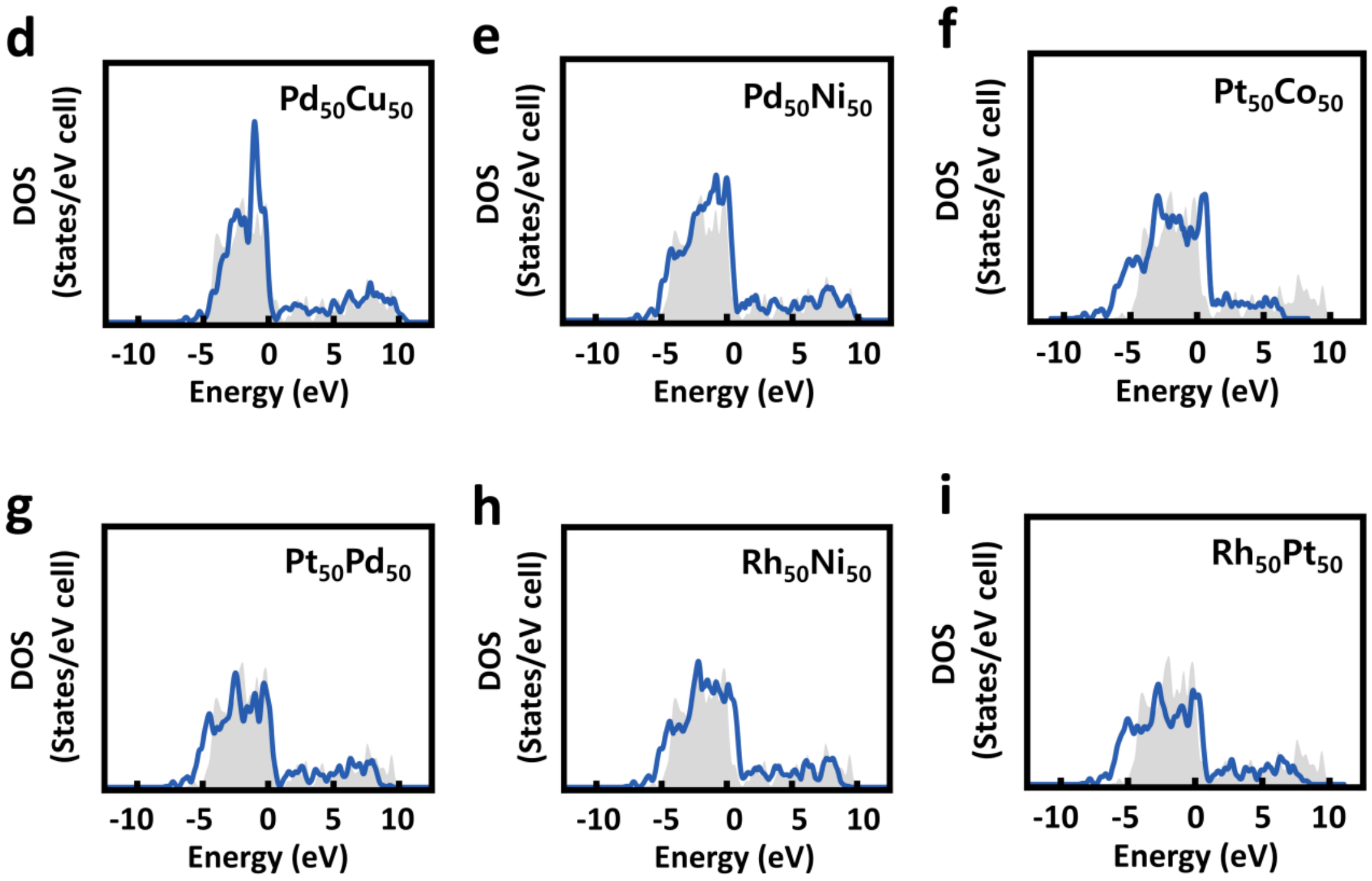

**Figure S11. Electronic DOS patterns of 9 bimetallic catalysts.** The blue lines indicate the DOS patterns of **a**. $Au_{50}Pd_{50}$(111), **b**. $Ni_{50}Co_{50}$(111), **c**. $Ni_{50}Pt_{50}$(111), **d**. $Pd_{50}Cu_{50}$(110), **e**. $Pd_{50}Ni_{50}$(111), **f**. $Pt_{50}Co_{50}$(111), **g**. $Pt_{50}Pd_{50}$(111), **h**. $Rh_{50}Ni_{50}$(111), and **i**. $Rh_{50}Pt_{50}$(111). The Fermi energy corresponds to the zero energy level. For comparison, the DOS pattern for Pd(111) (gray filled area) is included in each figure.

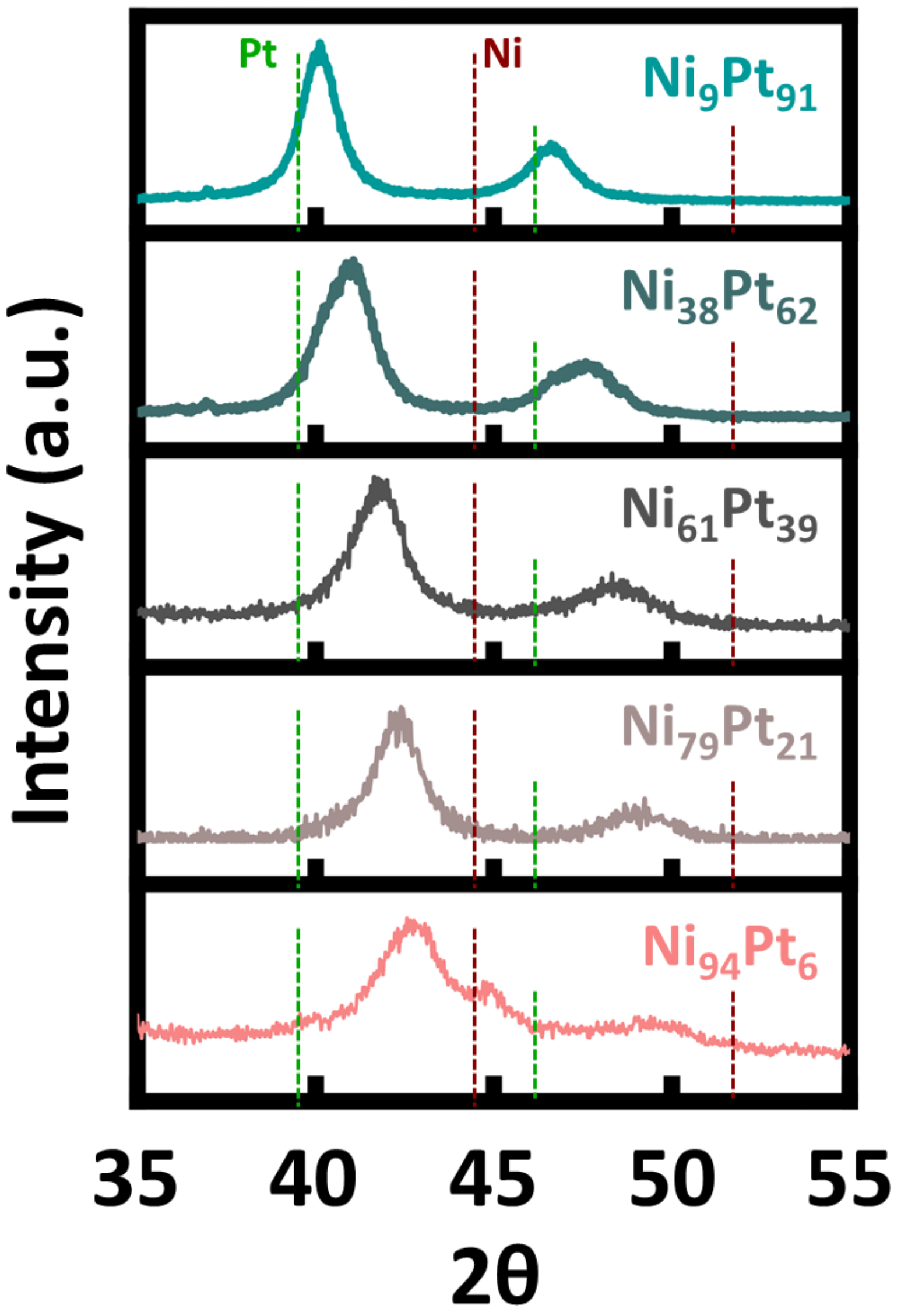

**Figure S12. XRDs of $Ni_xPt_{100-x}$ NPs.** No significant structural change in the NixPt100-x NPs was observed, although there is a small amount of Ni that was not alloyed with Pt in $Ni_{94}Pt_6$. For comparison, the XRD peaks for pure Ni (brown dotted line) and Pt (green dotted line) crystals are included.

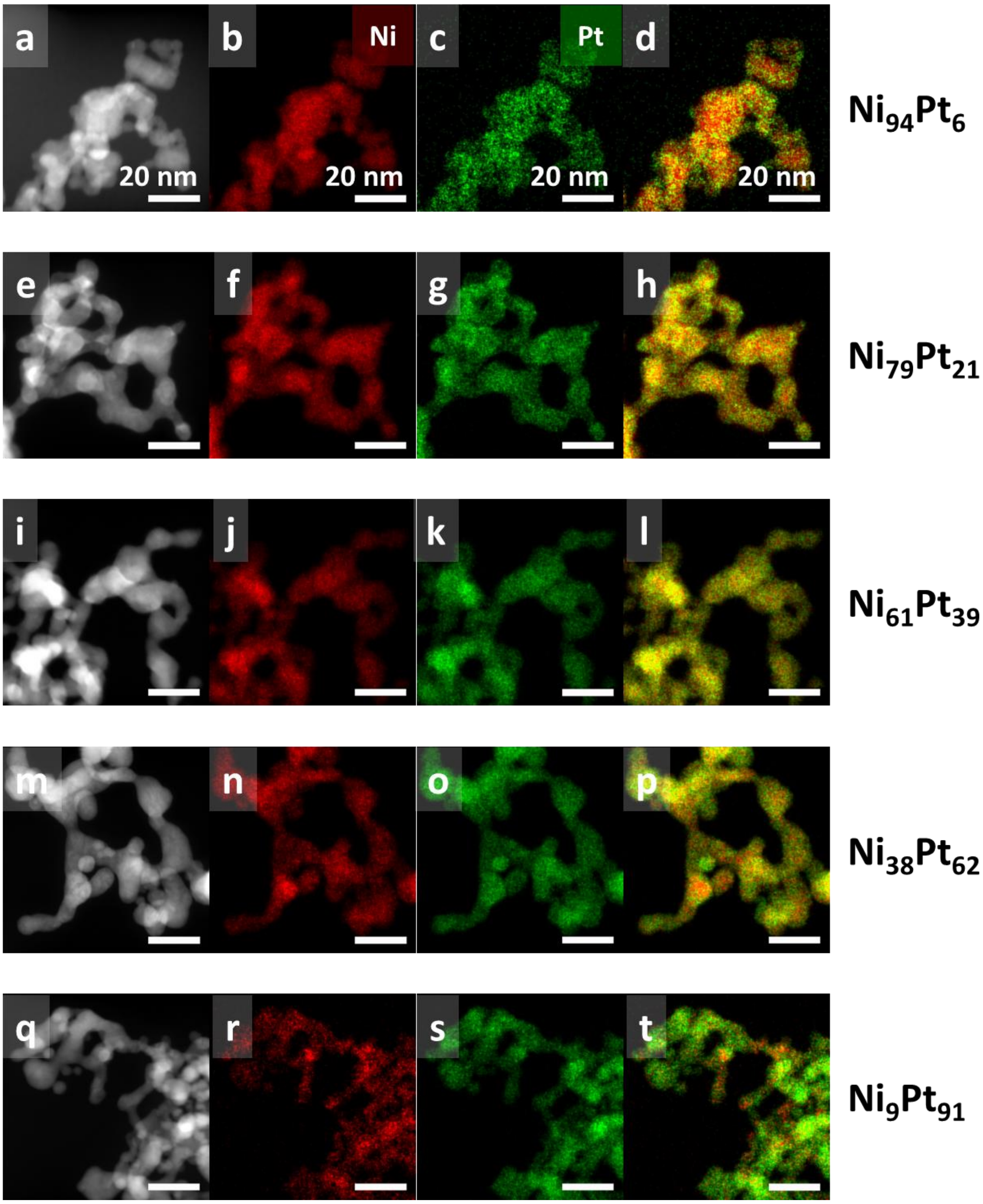

**Figure S13. Structural analysis of $Ni_xPt_{100-x}$ NPs. a-d**. $Ni_{94}Pt_6$, **e-h**. $Ni_{79}Pt_{21}$, **i-l**. $Ni_{61}Pt_{39}$, **m-p**. $Ni_{38}Pt_{62}$, and **q-t**. $Ni_9Pt_{91}$. HAADF-STEM images (**a**, **e**, **i**, **m**, and **q**), Ni-K STEM-EDX map (**b**, **f**, **j**, **n**, and **r**), Pt-L STEM-EDX map (**c**, **g**, **k**, **o**, and **s**), and the reconstructed Ni/Pt overlay image of the maps (**d**, **h**, **l**, **p**, and **t**). The scale bar in the images represents 20 nm.

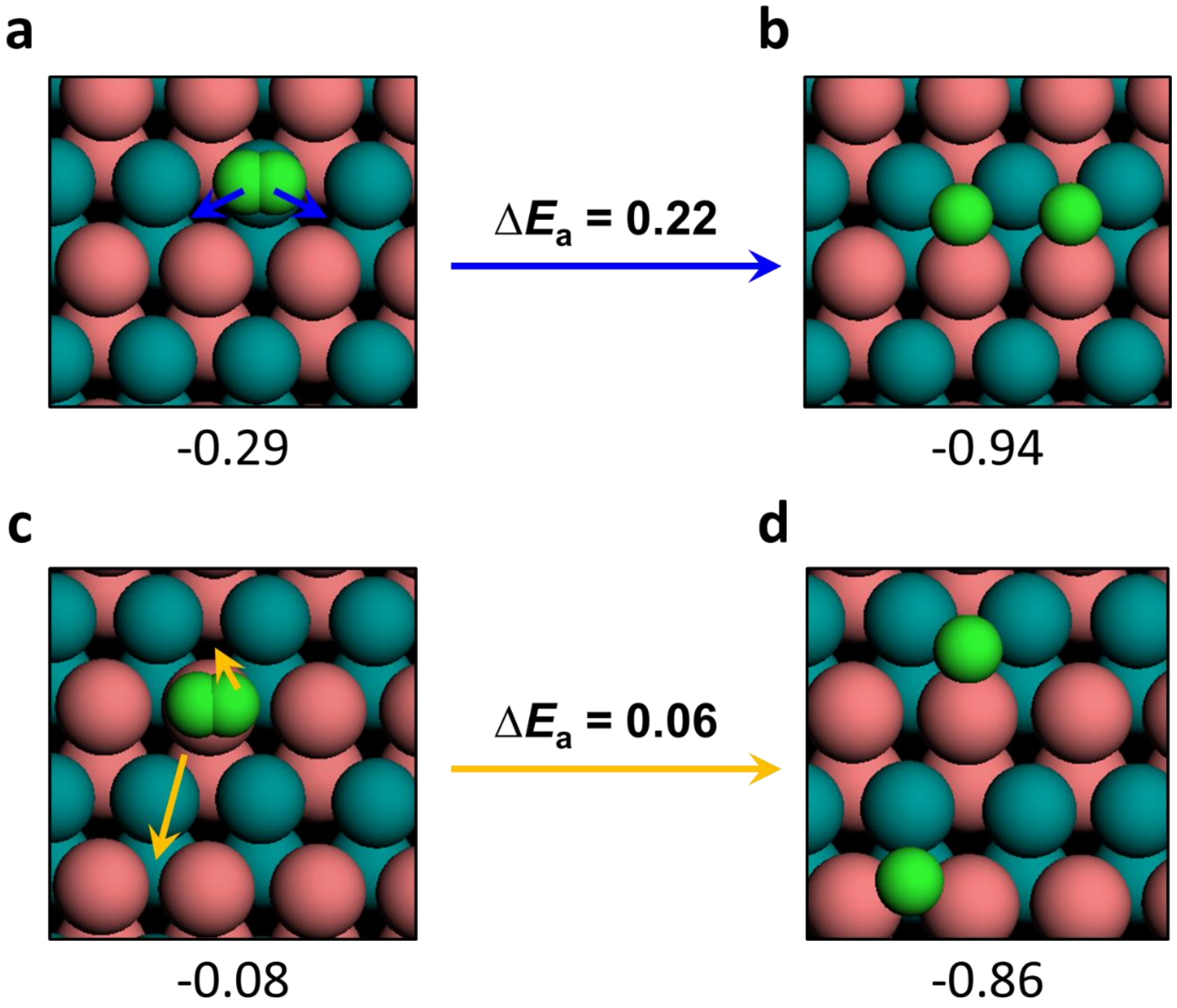

**Figure S14. DFT calculation of $H_2$ dissociation over $Ni_{50}Pt_{50}$(111). a**. Associative $H_2$ adsorption on the top Ni site, **b**. dissociative $H_2$ adsorption on two face-centered cubic (fcc) sites, **c**. associative $H_2$ adsorption on the top Pt site, and **d**. dissociative $H_2$ adsorption on 1 hcp and 1 fcc sites. The number below each figure corresponds to the binding energy in eV relative to the $Ni_{50}Pt_{50}$(111) slab and a $H_2$ molecule, where the negative value means a thermodynamically favorable state. The energy barriers ($\Delta E_a$) for $H_2$ dissociation are also included. The $H_2$ dissociation reactions were chosen after all possible associative and dissociative $H_2$ configurations were considered. The color code for the atoms is as follows: teal = Ni, coral = Pt, red = O, and green = H.